\documentclass[a4paper,superscriptaddress,twocolumn,showkeys,prX]{revtex4-1}
\usepackage{amssymb, amsmath}
\usepackage{graphics}
\usepackage{pst-all}
\usepackage{hyperref}
\usepackage[T1]{fontenc}
\usepackage[utf8]{inputenc}
\usepackage{epsfig}
\newcommand{\ud}{\,\mathrm{d}}

\begin{document}
\title{Phase transitions in the two-dimensional single-ion anisotropic ferromagnetic with long-range interactions}
\author{A. R. Moura}
\email{armoura@infis.ufu.br}
\affiliation{Universidade Federal de Uberlândia, Minas Gerais, Brazil}
\date{\today}

\begin{abstract}
In the present work, we investigate the effects of long-range interactions on the phase transitions of two-dimensional
ferromagnetic models with single-ion anisotropy at zero and finite temperatures. The Hamiltonian is given by
$H=\sum_{i\neq j} J_{ij}(S_i^xS_j^x+S_i^yS_j^y+\lambda S_i^zS_j^z)+D\sum_{i}(S_i^z)^2$, where
$J_{ij}=-J |r_j-r_i|^{-p}$ ($p\geq 3$) is a long-range ferromagnetic interaction ($J>0$) , $0\leq \lambda\leq 1$ is an
anisotropic constant and $D$ is the single-ion anisotropic constant. It is well-known that the single-ion anisotropy $D$
creates a competition between an ordered state (favored by the exchange interaction) and a disordered state, even at zero
temperature. For small values of $D$, the system has a spontaneous magnetization $m_z\neq 0$, while in the large-D
phase $m_z=0$ because a state with $\langle S^z\rangle\neq 0$ is energetically unfavorable. Therefore, a phase transition
due to quantum fluctuations occurs in some critical value $D_c$. For systems with short-range interaction $D_c\approx 6J$,
depending of $\lambda$ constant, but in our model we have found larger values of $D$ due to the higher cost to flip a spin.
Since low-dimensional magnetic systems with long range interaction can be ordered at finite temperature, we also have
analyzed the thermal phase transitions (similar to the BKT transition). The model has been studied by using a Schwinger boson
formalism as well as the Self-consistent Harmonic Approximation (SCHA) and both methods provide according results.
\end{abstract}

\keywords{single-ion anisotropic; long-range interaction; phase transitions; SCHA; Schwinger bosons}

\maketitle

\section{Introduction}
Usually, condensed matter physics considers short-range interactions between the nearest-neighbor spins as the main
responsible for the properties of low-dimensional magnetic systems, existing an extensive number of works on this topic.
However, longer-range interactions are also important to describe many other cases, where only short-range interactions are insufficient.
For example, in antiferromagnetic models, second-neighbor interaction between spins are responsible for an increase in
disorder due to lattice frustration once it is impossible a complete antiferromagnetic align. In addition, the properties of spin ice
systems arise as consequence of the long-range dipolar interaction. It creates a power-law decaying interaction between
the magnetic monopoles quasiparticles, similar to the electric charges. 

It is well known that low-dimensional systems are more susceptible to quantum fluctuations which imply no continuous
broken symmetry at finite temperatures for one and two-dimensional magnetic models with only short-range interactions
(Mermin-Wagner theorem\cite{PRL17}). On the other hand, the long-range interactions are able to create a state with
spontaneous magnetization at finite temperatures once the spin-waves do not have sufficient energy to flip spins. For a
$d$-dimensional Heisenberg model with interaction between the sites $i$ and $j$ decaying as $r^p$, there is a state with
spontaneous magnetization at finite low temperatures if $d<p<2d$ \cite{JP2,CMP62} while for $p\geq 2d$ the long-range
order ($LRO$) is lost at all finite temperatures \cite{JSP25,CMP79}. The condition $p>d$ is applied in order to avoid an energy
divergence per site in the ground state. If the symmetry is discrete instead continuous, as occurs in the Ising model,
there is $LRO$ for $1<p\leq 2$ \cite{CMP118}.

In this paper we have investigated the two-dimensional ($2d$) ferromagnetic model with long-range interactions and single-ion
anisotropy. The Hamiltonian is given by:
\begin{equation}
\label{eq:hamiltonian}
H=\sum_{i\neq j}J_{ij}\left(S_i^x S_j^x+S_i^y S_j^y + \lambda S_i^z S_j^z\right) + D\sum_i (S_i^z)^2,
\end{equation}
where $J_{ij}=-J|\vec{r}_j-\vec{r}_i|^{-p}$ ($J>0$) is the long-range ferromagnetic interaction, $\lambda$ is an anisotropic
constant and $D$ is the single-ion anisotropic coupling. The first sum is evaluated over all pairs of $i$ and $j$ sites while the
second one is over all sites. For $\lambda=0$ one has the usual $XY$ model (considering ${\bf S}=(S^x,S^y,S^z)$) while $\lambda=1$
gives the isotropic Heisenberg model. The Hamiltonian makes sense only for spins larger than $1/2$ once the single-ion
anisotropy is degenerate with respect the up and down states. Therefore, we have adopted $S=1$ for which there are two
energy bands: $S_i^z=\pm1$ and $S_i^z=0$. Though the finite energy ground state requires $p>2$, we have considered only
the decays with $p\geq 3$ because there is no physical models of interest with $2<p<3$. The one-dimensional case has been
studied by Pires \cite{PLA202} and a first-order thermal phase transition has been found for $XY$ model and Planar Rotator Model
(when ${\bf S}=({S^x,S^y)}$). The critical temperatures (the point where the spin-spin correlation vanishes) are consistent with
Monte Carlo simulations \cite{NCD10} for $p=3/2$ and $p=2$. However, for $2d$ models, only few results
are currently known, motivating us to develop this work.

In addition to the thermal phase transition, Hamiltonian (\ref{eq:hamiltonian}) also presents a quantum phase transition ($QFT$)
associated with the single-ion anisotropy. There is a competition between the exchange and the single-ion anisotropic energy.
For small values of $D$, the quantum fluctuations are negligible and the spins tend to remain align with the $z$-axis, providing
$\langle S^z \rangle\neq 0$. On the other hand, in the large-$D$ phase, the quantum fluctuations are larger and the energetic cost to keep
a non-null $S^z$ spin component is too high. Therefore, the system goes to a gap phase with no spontaneous magnetization
even at zero temperature. This transition occurs at the point $D=D_c$ and it is well documented for models with short-range
interaction \cite{JPCM5, PRB48, PRB49, PRB66, PRB71, JPCM20}. Below the critical point $D_c$, the system has a gapless energy
spectrum and a quasi long-range order (quasi-$LRO$) with algebraic decay for the spin order-parameter correlation at low finite
temperature. Above a critical transition temperature $T_c$, the order-parameter decays exponentially and there is no more
quasi-$LRO$. Applying two different analytical methods, we have studied the quantum and thermal phase transitions as a function
of the power-law exponent. In the next section we apply the bond operator (a Schwinger bosonic formalism) to determine the
quantum phase transition at zero temperature. In section (\ref{scha}), we have developed the Self-consistent Harmonic Approximation
($SCHA$) at zero temperature and in section (\ref{finiteT}) the $SCHA$ is applied at finite temperatures. The conclusions are present
in section (\ref{conclusion}).

\section{Bond operator formalism}
\label{BO}
In this section, we use a $SU(3)$ Schwinger bosonic formalism, the so-called bond operator, to describe the $QFT$ at zero
temperature in the large-$D$ phase. As commented earlier, for $D>D_c$ the system is disordered (vanishing spin-spin correlation) at
zero temperature due to spin-waves fluctuations. To order to determine the critical point $D_c$, we have followed the procedures
of references \cite{PRB56,PRB71,PLA360}. Considering a spin-$1$ model, we define three bosonic operators to represent the
states of $S^z$: $|m_i=-1\rangle=a_{i,-1}^\dagger|0\rangle$, $|m_i=0\rangle=a_{i,0}^{\dagger}|0\rangle$ and
$|m_i=1\rangle=a_{i,1}^\dagger|0\rangle$, where $|0\rangle$ is the vacuum state of the Fock space. The boson operator
$a^\dagger_{im}$ creates a particle with $m_z=m$ on site $i$. The commutation relations $[S_i^+,S_j^-]=2S_i^z\delta_{ij}$ and
$[S_i^z,S_j^{\pm}]=\pm S_i^{\pm}\delta_{ij}$ are valid and, to keep $S^2_i=S(S+1)$, we have to impose a local constraint
$\sum_\mu a^\dagger_{i,\mu} a_{i,\mu}=S$ on each site. A condensation occurs in the $|m=0\rangle$ state once this is the smaller
band energy (the $|m=\pm 1\rangle$ excited states are degenerate). Therefore, $N_0$, the number of particles in $|m=0\rangle$
state, obeys the condition $N_0=\langle a^\dagger_{i,0} a_{i,0}\rangle\gg 1$ and we can consider the approximation $[N_0,a^\dagger_{i,0}]=0$.
Once $N_0$ and $a_{i,0}$ commute, we have treated the $a_{i,0}$ quantum operators as classical entities, i.e. real numbers.
Thus we have adopted the mean-values $\langle a^\dagger_{i,0}\rangle=\langle a_{i,0}\rangle=\rho_0^{1/2}$  in the next equations
($\rho_0$ measures the density condensate in state $a_0^\dagger|0\rangle$). In the $SU(3)$ bosonic representation, the
spin operators are written as:
\begin{eqnarray}
S_i^+&=&\sqrt{2\rho_0}(u_i^\dagger+d_i),\nonumber\\
S_i^-&=&\sqrt{2\rho_0}(d_i^\dagger+u_i),\nonumber\\
S_i^z&=&u_i^\dagger u_i-d_i^\dagger d_i,
\end{eqnarray}
on which we have properly replaced $a_{i,1}$ and $a_{i,-1}$ by $u_i$ and $d_i$, respectively. Therefore:
\begin{eqnarray}
&&S_i^xS_j^x+S_i^yS_j^y+\lambda S_i^zS_j^z=\rho_0(d_id_j^\dagger+ d_iu_j + u_i^\dagger d_j^\dagger + \nonumber\\
&&+ u_i^\dagger u_j+d_i^\dagger d_j + d_i^\dagger u_j^\dagger + u_id_j+u_i u_{j}^\dagger)\nonumber\\
&& +\lambda(u_i^\dagger u_i-d_i^\dagger d_i)(u_j^\dagger u_j-d_j^\dagger d_j).
\end{eqnarray}
The four operator terms are decoupled using a Hubbard-Stratonovich transform:
\begin{eqnarray}
&&(u_i^\dagger u_i-d_i^\dagger d_i)(u_j^\dagger u_j-d_j^\dagger d_j)=2p^2-\frac{1}{2}(1-\rho_0)^2-\nonumber\\
&&-\frac{1}{2}m^2+\frac{1}{2}(1-\rho_0+m)(u_i^\dagger u_i+u_j^\dagger u_j)+\nonumber\\
&&+\frac{1}{2}(1-\rho_0-m)(d_i^\dagger d_i+d_j^\dagger d_j)-p(u_i d_j+d_iu_j+\nonumber\\
&&+d_j^\dagger u_j^\dagger+u_j^\dagger d_i^\dagger),
\end{eqnarray}
where we have introduced  the mean-field parameters $m=\langle u_i^\dagger u_i\rangle-\langle d_i^\dagger d_i\rangle$
and $p=\langle d_i^\dagger u_j^\dagger\rangle=\langle d_i u_j\rangle$. The anisotropic term is written as
$(S_i^z)^2=u_i^\dagger u_i+d_i^\dagger d_i$ and after a Fourier transform, the Hamiltonian (\ref{eq:hamiltonian}) is given by:
\begin{widetext}
\begin{eqnarray}
\label{eq:hamiltonian_k}
H&=&\frac{1}{2}\sum_{\bf k}\left\{Z\left[\rho_0 J_{\bf k}-\lambda J_{\bf k} p\right](u_{-{\bf k}}d_{\bf k}+d_{-{\bf k}}u_{\bf k})+\left[\rho_0 ZJ_{\bf k}+\frac{ZJ_{\bf 0}}{2}(1-\rho_0+m)\lambda+D-\mu\right]u_{\bf k}^\dagger u_{\bf k}+\right.\nonumber\\
&&+\left.\left[\rho_0 ZJ_{\bf k}+\frac{ZJ_{\bf 0}}{2}(1-\rho_0 -m)\lambda+D-\mu\right]d_{\bf k}^\dagger d_{\bf k}+\textrm{H.c.}\right\}+H_0,
\end{eqnarray}
\end{widetext}
with $H_0=\mu N (1-\rho_0)+\frac{\lambda}{4} NZ\left[4p^2-m^2-(1-\rho_0)^2\right]$, $Z=N-1$ and $N$ is the lattice sites
number. In the last equation the constraint $\sum_\mu a^\dagger_{i,\mu} a_{i,\mu}=S$ was introduced as a Lagrange
multiplier $\mu$ and we have defined
\begin{eqnarray}
ZJ_{\bf k}=-JS_p({\bf k})=-J\sum_{\Delta{\bf r}\neq {\bf 0}} \frac{e^{-i {\bf k}\Delta{\bf r}}}{\Delta r^p}
\end{eqnarray}
as the Fourier transform of the long-range ferromagnetic interaction. The function $S_p({\bf k})$ is calculated using the
Ewald's method, where the sum is divided in two rapidly convergent sums: the first over the space lattice and the second over the
reciprocal lattice \cite{PR99,JPC14}. After a straightforward calculation, we obtain for a square two-dimensional lattice \cite{PRB22}:
\begin{eqnarray}
S_p({\bf k})&=&a^{-p}\frac{(\pi/4)^{p/2}}{\Gamma(p/2)}\left[\sum_{\bf l}e^{-ia{\bf k}{\bf l}}\Phi_{p/2-1}\left(\frac{\pi l^2}{4}\right)+\right.\nonumber\\
&&\left.+4\sum_{\bf l}\Phi_{-p/2}\left(\frac{|a{\bf k}-2\pi{\bf l}|^2}{\pi}\right)-\frac{2}{p}\right],
\end{eqnarray}
where $\Delta{\bf r}=a{\bf l}=a(l_1\hat{x}+l_2\hat{y})$, $a$ being the lattice parameter and $l_1$ and $l_2$ integers numbers.
The $\Phi$ functions are given by:
\begin{eqnarray}
\Phi_m(x)=\int_1^\infty y^m e^{-y x}dy.
\end{eqnarray}
The Hamiltonian (\ref{eq:hamiltonian_k}) is diagonalized by a Bogoliubov transform which provides:
\begin{eqnarray}
H-E_g=\sum_{\bf k}\left[\omega_{\bf k}^{(+)}\alpha_{\bf k}^\dagger\alpha_{\bf k}+\omega_{\bf k}^{(-)}\beta_{\bf k}^\dagger\beta_{\bf k}\right],
\end{eqnarray}
with
\begin{eqnarray}
\omega_{\bf k}^{(\pm)}=\omega_{\bf k} \pm\frac{1}{2}Z\lambda m,
\end{eqnarray}
\begin{eqnarray}
\omega_{\bf k}=\sqrt{\Lambda_{\bf k}^2-\Xi_{\bf k}^2},
\end{eqnarray}
\begin{eqnarray}
\Lambda_{\bf k}=D-\mu +\frac{1}{2}ZJ_{\bf 0}\lambda(1-\rho_0)+\rho_0 ZJ_{\bf k}
\end{eqnarray}
and
\begin{eqnarray}
\Xi_{\bf k}=(\rho_0-\lambda p)ZJ_{\bf k}.
\end{eqnarray}
The ground state energy is $E_g=\sum_{\bf k} \left(\omega_{\bf k}-\Lambda_{\bf k}\right) +H_0$. The new and old bosonic
operators are related by:
\begin{eqnarray}
d_{\bf k}=\xi_{\bf k}\beta_{\bf k}-\zeta_{\bf k}\alpha_{-{\bf k}}^\dagger
\end{eqnarray}
and
\begin{eqnarray}
u_{\bf k}=\xi_{\bf k}\alpha_{\bf k}-\zeta_{\bf k}\beta_{-{\bf k}}^\dagger
\end{eqnarray}
where $\xi_{\bf k}^2=\frac{1}{2}(1+\frac{\Lambda_{\bf k}}{\omega_{\bf k}})$ and $\zeta_{\bf k}^2=\frac{1}{2}(-1+\frac{\Lambda_{\bf k}}{\omega_{\bf k}})$.
The mean-field parameters $\rho_0$, $m$, $p$ and $\mu$ are determined  by the the self-consistent equations obtained
from the minimum of the Helmoltz free energy $F=-\beta^{-1}\ln\textrm{Tr}( e^{-\beta H})$. At zero temperature, the
continuous limit of the self-consistent equations is given by:
\begin{eqnarray}
\label{eq:rho}
\rho_0=2-\frac{1}{4\pi^2}\int \frac{\Lambda_{\bf k}}{\omega_{\bf k}}\ud^2{\bf k},
\end{eqnarray}
\begin{eqnarray}
\label{eq:p}
p=-\frac{1}{8\pi^2}\int\frac{\Xi_{\bf k}(J_{\bf k}/J_{\bf 0})}{\omega_{\bf k}}\ud^2{\bf k},
\end{eqnarray}
and
\begin{eqnarray}
\label{eq:mu}
\mu=\frac{1}{4\pi^2}\int\frac{(\Lambda_{\bf k}-\Xi_{\bf k})ZJ_{\bf k}}{\omega_{\bf k}}\ud^2{\bf k},
\end{eqnarray}
where the integrals are evaluated over the first Brioullin Zone. The magnetization parameter $m$ is identically null at zero
temperature (in the large-$D$ phase) and we can also approach $p\approx0$.

\begin{figure}[h]
\centering
\epsfig{file=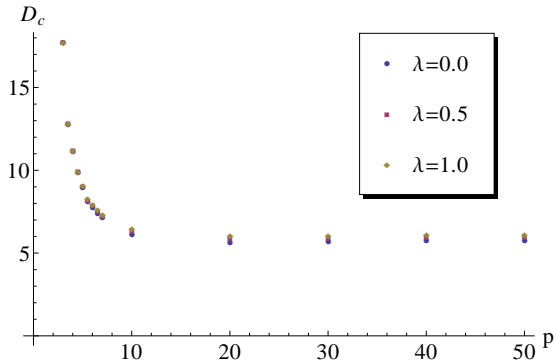,width=\linewidth}
\caption{The critical value $D_c$ (in units of $J$) as function of the power-law exponent $p$.}
\label{fig:dc_bo}
\end{figure}

The energy $\omega_{\bf k}$ has a minimum at ${\bf k}^\ast=(0,0)$ and a finite gap $\Delta=\omega_{{\bf k}^\ast}$ exist for
$D>D_c$. Even at zero temperature there is no spontaneous magnetization once the bosons condensate in the
$|m=0\rangle$ state. Close to the critical point, $D\gtrsim D_c$, the gap decreases and when $D=D_c$, a quantum phase
transition occurs. Below $D_c$, a gapless state takes place, which provides $\langle S^z\rangle\neq 0$, because the
states $|m=1\rangle$ (or $|m=-1\rangle$) are filled. The bosonic formalism developed here is applied only for the large-$D$
phase. A similar method for $D<D_c$ phase can also be formulated, although, to our proposals, it is not necessary.
Defining $y=2\rho_0\left[D-\mu+\frac{\lambda ZJ_{\bf 0}}{2}(1-\rho_0)\right]^{-1}$, equations (\ref{eq:rho}) and (\ref{eq:mu})
can be rewritten as:
\begin{eqnarray}
2(2-\rho_0)=\left[I_1(y_m) + I_2(y_m)\right]
\end{eqnarray}
and
\begin{eqnarray}
y\mu=I_2(y_m)-I_1(y_m),
\end{eqnarray}
which $I_1(y_m)=\frac{1}{4\pi^2}\int\frac{\ud^2{\bf k}}{\sqrt{1+y ZJ_{\bf k}}}$ and
$I_2(y_m)=\frac{1}{4\pi^2}\int\sqrt{1+yZJ_{\bf k}}\ud^2{\bf k}$. In the gapless point ($D=D_c$) we have the minimum
$y_m=-\frac{1}{ZJ_{\bf 0}}$ and using equations above, the critical point is found as:
\begin{eqnarray}
D_c&=&\frac{\lambda ZJ_{\bf 0}}{4}\left[2-I_1(y_m)-I_2(y_m)\right]+\nonumber\\
&&+\frac{2}{y_m}\left[2-I_1(y_m)\right].
\end{eqnarray}
The results for $D_c$ as function of the power-law exponent $p$ are show in Fig. (\ref{fig:dc_bo}). The $\lambda$ influence
is small, mainly for $p<10$, and both $XY$ and isotropic model present approximately the same critical anisotropic constant
$D_c$. For small $p$, the larger critical point ($D_c=17.72 J$ if $p=3$ and $D_c=12.86 J$ when $p=4$, both with $\lambda=1$)
reflects the higher energetic cost to flip spins. Due the strong long-range interaction, it is easer to keep an ordered state which
minimizes the exchange energy ($S_z\neq 0$) instead a state with $S_z=0$. To invert the dispute, making more favorable a
state which minimizes the single-ion anisotropic energy, it is necessary a large anisotropic constant $D$. On the other hand,
for $p\gg 1$ the $D_c$ tends to a constant value, $D_c=5.74 J$ if $\lambda=0$ and $D_c=6.10 J$ when $\lambda=1$.
As expected, for long distances $\Delta r_{ij}$, the interaction is negligible and the system behaves like a model with short-range
interactions. Therefore, the results for large $p$ agree with that obtained from models with nearest neighbors interactions. For two-dimensional
antiferromagnetic models ($J=-1$) with only $SR$ interactions, we have $D_c=5.72 J$ for the $XY$ model \cite{PA388} and the set
of values $D_c=5.77 J$ \cite{PLA360}, $D_c=5.82 J$ \cite{PRB71} and $D_c=6.38 J$ \cite{PRB50} for the isotropic model.
The same values are obtained also for a ferromagnetic coupling ($J=1$), once the condensation occurs at ${\bf k}=(0,0)$ instead
${\bf k}=(\pi,\pi)$ (for the antiferromagnetic case) and the Hamiltonian is invariant under the transformation
${\bf k}\to{\bf k}+{\pi}$ and $J\to -J$.

\begin{figure}[h]
\centering
\epsfig{file=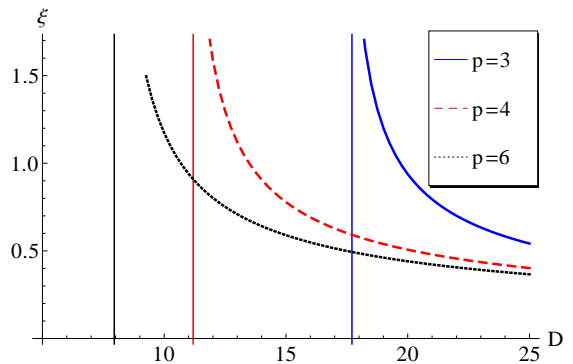,width=\linewidth}
\caption{The correlation length as function of the anisotropic constant $D$. The vertical lines indicate the critical points
$D_c$ for each $p$. The constants are given in units of $J$.}
\label{fig:xi_bo}
\end{figure}

Close to the origin we can adopt the approximation: $S_p(k)\approx S_p(0)-\kappa_p f_p(k)$ where
$S_p(0)=4^{1-n/2}\left[\zeta(\frac{p}{2},\frac{1}{4})-\zeta(\frac{p}{2},\frac{3}{4})\right]\zeta(\frac{p}{2})$, with $\zeta(s,q)$
being the Hurwitz zeta function and $\zeta(s)$ the zeta function. The constant $\kappa_p$ is numerically determined for
each exponent $p$ and the function $f_p(k)$ is defined as \cite{PRB46,PLA202}:
\begin{eqnarray}
f_p(k)=\left\{
\begin{array}{lc}
k^{p-2}, & 2<p<4\\
k^2 \ln(k), &  p=4\\
k^2, &  p>4
\end{array}
\right.
\end{eqnarray}
Thus, the excited state energy for $|{\bf k}-{\bf k}^\ast|\approx 0$ is $\omega_{\bf k}=\sqrt{\Delta^2+c^2 f_p^2(k)}$, where
the gap energy and the long wavelength spin-wave velocity are given by:
\begin{eqnarray}
\Delta=\left[D-\mu+\frac{\lambda ZJ_{\bf 0}}{2}(1-\rho_0)\right]\sqrt{1+ZJ_{\bf 0} y}
\end{eqnarray}
and
\begin{eqnarray}
c=\left[D-\mu+\frac{\lambda ZJ_{\bf 0}}{2}(1-\rho_0)\right]\sqrt{J \kappa_p y},
\end{eqnarray}
respectively . For power-law decay with exponent $p>4$ we have relativistic dispersion relation, as occurs in the short-range
interaction models, and the correlation length is expressed by:
\begin{eqnarray}
\label{eq:cl}
\xi=\frac{c}{\Delta}=\sqrt{\frac{J\kappa_p y}{1+ZJ_{\bf 0} y}}.
\end{eqnarray}
In the limit of low temperatures and long-distance $\Delta r_{ij}$, the equal time correlation function is given by:
\begin{eqnarray}
&& \langle S_i^+ S_j^- \rangle =\frac{1}{4\pi^2}\int\langle S_k^+ S_k^- \rangle e^{i{\bf k}\cdot\Delta{\bf r}}\ud^2{\bf k}\nonumber\\
&&=\frac{\rho_0}{\pi^2}\int\left(\frac{\Delta_{\bf k}-\Xi_{\bf k}}{\omega_{\bf k}}\right)\left(n_{\bf k}+\frac{1}{2}\right)e^{i{\bf k}\cdot\Delta{\bf r}}\ud^2{\bf k}\nonumber\\
&&\approx \frac{T}{2J \kappa_p \pi^2}\int\frac{e^{i{\bf k}\cdot\Delta{\bf r}}}{k^2+\xi^{-2}}\ud^2{\bf k} \propto \left(\frac{\Delta r}{\xi}\right)^{-\frac{1}{2}}e^{-\frac{\Delta r}{\xi}},
\end{eqnarray}
where $\xi$ is finite in the large-$D$ phase. As expected, even at zero temperature, there is no $LRO$ when $D>D_c$ and the
correlation function is similar to that obtained from the models with short-range interactions if $p>4$. When $D=D_c$ the system goes to a gapless
state which provides an infinite correlation function and $LRO$ for $D<D_c$. In Fig. (\ref{fig:xi_bo}), we plot the correlation length
at zero temperature for $p=2$, $4$ and $6$. In the large-$D$ phase, $\xi$  is finite but diverges when $D$ tends to $D_c$ (represented by the vertical lines).

The dynamical structure factor is calculated as:
\begin{eqnarray}
S({\bf k},t)&=&\langle S_k^+(t)S_k^-(0)\rangle \nonumber\\
&=& \langle S_k^+S_k^-\rangle\left[e^{-i\omega_{\bf k}t}H(t)-e^{i\omega_{\bf k}t}H(-t)\right],
\end{eqnarray}
where $H(t)$ is the Heaviside step function. Using $H(t)=\displaystyle\lim_{\epsilon\to 0^+}\frac{1}{2\pi i}\int_{-\infty}^{+\infty}\frac{e^{itx}}{x-i\epsilon}dx$,
the time Fourier transform provides:
\begin{eqnarray}
S({\bf k},\omega)=\frac{2c^2}{J\kappa_p(\omega^2-\omega_{\bf k}^2)}.
\end{eqnarray}

\section{Self-consistent Harmonic Approximation}
\label{scha}
The Self-consistent Harmonic Approximation ($SCHA$) \cite{JPC7,SSC104,PA388,PRB48,PRB53,PRB54,PSS242,JMMM357} is the most appropriate
spin-wave method to treat the model at finite temperatures. The bosonic formalism used in the last sections provides reasonable
results at zero temperature but, as argued by Yoshida \cite{JPSJ58}, for $T>0$ the Schwinger formalism presents some
divergences with another well-known results.

\begin{figure}[h]
\centering
\epsfig{file=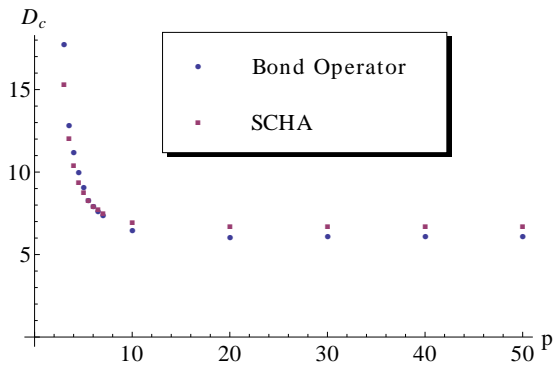,width=\linewidth}
\caption{The critical value $D_c$ (in units of $J$) as function of the power-law exponent $p$.}
\label{fig:dc_scha}
\end{figure}

As demonstrated, due to the single-ion anisotropy, the analyzed system exhibits a quantum phase transition at $D=D_c$. In
the large-$D$ phase, $D>D_c$, there is no spontaneous magnetization at finite and zero temperature. Below the critical point
$D_c$, the $O(3)$ symmetry is spontaneously broken and $m\neq 0$ at zero temperature. If $2<p<4$, there is also $LRO$ at finite
low temperatures (above a transition temperature the disordered state takes place). However, for $p\geq 4$ there is no
spontaneous broken symmetry at any finite temperature and the system behaves like a short-range model for which the
Mermin-Wagner theorem is valid. There is a thermal phase transition similar to the Berezinski-Kosterlitz-Thouless
transition present in the $XY$ model. Below a critical temperature $T_c$, the model has a quasi long-range order with algebraic
decay for the spin order-parameter correlation. For $T>T_c$ the decay becomes exponential and there is no more quasi
$LRO$. In this section we have applied the $SCHA$ to determine the critical temperature $T_c$ beyond the $D_c$ constant. The
main idea behind the $SCHA$ is replace the original Hamiltonian by another one with temperature dependent renormalized
parameters. Using the Villain representation \cite{JP35}, we written the spin operators as:
\begin{eqnarray}
S_i^+=e^{i\phi_i}\sqrt{S(S+1)-S_i^z(S_i^z+1)}
\end{eqnarray}
and
\begin{eqnarray}
S_i^-=\sqrt{S(S+1)-S_i^z(S_i^z+1)}e^{-i\phi_i}.
\end{eqnarray}
Therefore the Hamiltonian (\ref{eq:hamiltonian}) is given by:
\begin{widetext}
\begin{eqnarray}
\label{eq:hscha}
H=\frac{1}{2}\sum_{i\neq j}J_{ij}\left[\tilde{S}^2\sqrt{1-\left(\frac{S_i^z}{\tilde{S}}\right)^2}\sqrt{1-\left(\frac{S_j^z}{\tilde{S}}\right)^2}\cos(\phi_j-\phi_i)+
\lambda S_i^zS_j^z\right]+D\sum_i(S_i^z)^2,
\end{eqnarray}
\end{widetext}
with $\tilde{S}=S(S+1)$. The mean-value $\langle \phi\rangle$ is not well defined to angles measured relative to a fixed axis,
so we choose the angle operator $\phi_i$ relative to the direction of the instantaneous magnetization in order to avoid divergences.
Considering $D<D_c$ and low temperatures, the spin field assumes a configuration with a small angular difference,
$|\phi_j-\phi_i|\ll 1$. Thus, we expand the Hamiltonian in powers of $(S_i^z/\tilde{S})^2$ and $(\phi_j-\phi_i)^2$ which provides:
\begin{eqnarray}
\label{eq:hscha0}
H&=&\frac{1}{2}\sum_{i\neq j}J_{ij}\left[-\frac{\rho_s \tilde{S}^2}{2}(\phi_j-\phi_i)^2-(S_i^z)^2+\lambda S_i^zS_j^z\right]+\nonumber\\
&&+D\sum_i (S_i^z)^2,
\end{eqnarray}
where the spin stiffness:
\begin{eqnarray}
\rho_s=\left\langle\sqrt{1-\left(\frac{S_i^z}{\tilde{S}}\right)^2}\sqrt{1-\left(\frac{S_j^z}{\tilde{S}}\right)^2}\cos(\phi_j-\phi_i)\right\rangle.
\end{eqnarray}
takes into account contributions of anharmonic terms in the approximation. The average $\langle\cdots\rangle$
is taken with respect to the original Hamiltonian (\ref{eq:hscha}). Once the fields $\phi_i$ and $S_i^z$ obey the
fundamental Poisson bracket $\{\phi_i,S_i^z\}=\delta_{ij}$ for polar representation of a spin vector, the operators $\phi_i$
and $S_i^z$ are canonically conjugate, i.e. $[\phi_i,S_j^z]=i\delta_{ij}$ (adopting $\hbar=1$). After a Fourier transform, we have:
\begin{eqnarray}
\label{eq:hschak}
H&=&\frac{1}{2}\sum_{\bf k}\left\{\left[ Z(\lambda J_{\bf k}-J_{\bf 0})+2D\right]S_{\bf k}^zS_{-{\bf k}}^z\right.+\nonumber\\
&&\left.+\rho_s\tilde{S}^2Z(J_{\bf k}-J_{\bf 0})\phi_{\bf k}\phi_{-{\bf k}}\right\},
\end{eqnarray}
on which $J_{\bf k}$ is the same defined in previous section. The Hamiltonian is then diagonalized introducing a new bosonic
operator $a_{\bf k}$ by the canonical transformation:
\begin{eqnarray}
\phi_{\bf k}=\frac{1}{\sqrt{2}}\left[\frac{Z(\lambda J_{\bf k}-J_{\bf 0})+2D}{\rho_s \tilde{S}^2Z(J_{\bf k}-J_{\bf 0})}\right]^{\frac{1}{4}}(a_{\bf k}^\dagger+a_{-{\bf k}}),
\end{eqnarray}
and
\begin{eqnarray}
S_{\bf k}^z=\frac{i}{\sqrt{2}}\left[\frac{\rho_s \tilde{S}^2Z(J_{\bf k}-J_{\bf 0})}{Z(\lambda J_{\bf k}-J_{\bf 0})+2D}\right]^{\frac{1}{4}}(a_{\bf k}^\dagger-a_{-{\bf k}}).
\end{eqnarray}
After a straightforward calculation, we obtain the harmonic diagonalized Hamiltonian:
\begin{eqnarray}
H_0=\sum_{\bf k}\omega_{\bf k}\left(a_{\bf k}^\dagger a_{\bf k} +\frac{1}{2}\right),
\end{eqnarray}
with the spin-wave energy
\begin{eqnarray}
\omega_{\bf k}=\sqrt{\rho_s\tilde{S}^2Z(J_{\bf k}-J_{\bf 0})[Z(\lambda J_{\bf k}-J_{\bf 0})+2D]}.
\end{eqnarray}
Through the diagonalized Hamiltonian, we have determined the mean-value fields:
\begin{eqnarray}
\langle\phi_{\bf k}\phi_{-{\bf k}}\rangle_0=\frac{1}{2}\sqrt{\frac{Z(\lambda J_{\bf k}-J_{\bf 0})+2D}{\rho_s \tilde{S}^2Z(J_{\bf k}-J_{\bf 0})}}\coth\left(\frac{\beta\omega_{\bf  k}}{2}\right)
\end{eqnarray}
and
\begin{eqnarray}
\langle S_{\bf k}^zS_{-{\bf k}}^z\rangle_0=\frac{1}{2}\sqrt{\frac{\rho_s\tilde{S}^2Z(J_{\bf k}-J_{\bf 0})}{Z(\lambda J_{\bf k}-J_{\bf 0})+2D}}\coth\left(\frac{\beta\omega_{\bf  k}}{2}\right)
\end{eqnarray}
To evaluate the spin stiffness, we calculate the average $\langle\cdots\rangle$ using the diagonalized Hamiltonian $H_0$. Since
$\phi_i$ and $S_i^z$ are uncoupled operators and $\phi$ has a Gaussian distribution, $\rho_s$ is given by:
\begin{eqnarray}
\label{eq:rho_scha}
\rho_s\approx\left[1-\left\langle\left(\frac{S_i^z}{\tilde{S}}\right)^2\right\rangle_0\right]e^{-\frac{1}{2}\langle(\phi_j-\phi_i)^2\rangle_0}.
\end{eqnarray}
For small values of $D$, the spin-waves have small energy and the spin stiffness is finite; however, in the large-$D$ phase, the
quantum fluctuations are sufficiently large, disordering the system. The point (at zero temperature) where $\rho_s$ abruptly
vanishes is taken as the critical point $D_c$. Using Eq. (\ref{eq:rho_scha}), as well as the the mean-values for
$\langle\phi_{\bf k}\phi_{-{\bf k}}\rangle_0$ and  $\langle S_{\bf k}^z S_{-{\bf k}}^z\rangle_0$, we have determined the transition
point $D_c$ as a function of the power-law exponent $p$. In Fig. (\ref{fig:dc_scha}) we plot the results from $SCHA$ analysis (for $\lambda=1$)
and also those obtained in a previous section (using the bond operator formalism). As one can see, both methods provide according
results. In the limit $p\gg 1$, in which the interactions behave like a short-range, the critical anisotropy is $D_c=6.72J$, slightly larger
than those obtained from bond operator method ($D_c=6.10J$).

\section{Finite temperature}
\label{finiteT}
As commented previously, beyond the quantum phase transition at $D=D_c$, there is a thermal transition which separates a phase
with quasi-long-range order from another with null spin-spin correlation. Applying Eq. (\ref{eq:rho_scha}) to finite temperatures we
calculate the critical point $T_c$ as the point where $\rho_s$ vanishes discontinuously. This temperature, as explained in next
section, is a first approximation to a transition of $BKT$ kind. In the classical limit, we can use the approximation
$\coth\left(\frac{\beta\omega_{\bf k}}{2}\right)\approx \frac{2T}{\omega_{\bf k}}$ (adopting $k_B=1$). Equation (\ref{eq:rho_scha}) is
then written as:
\begin{eqnarray}
\label{eq:rho_cl}
\rho_s=(1-I t)e^{-\frac{R t}{\rho_s}},
\end{eqnarray}
in which $t=\frac{T}{J \tilde{S}^2}$ is the reduced temperature and the constants are given by:
\begin{eqnarray}
R=\frac{1}{4\pi}\int \frac{J(1-\gamma_{\bf k})}{Z(J_{\bf k}-J_{\bf 0})}\ud^2{\bf k}
\end{eqnarray}
and
\begin{eqnarray}
I=\frac{1}{4\pi}\int \frac{J}{Z(\lambda J_{\bf k}-J_{\bf 0})+2D}\ud^2{\bf k}.
\end{eqnarray}
The $I$ term measures the out-of-plane fluctuations and therefore $I=0$ for the planar rotator model since no $S^z$ components
are allowed. For the other models $0<I\leq 1$. It is important to highlight that better results are reached for large $p$ since
this case is closer to a short-range interaction model. For small power-law exponents $p$, the energy of the spin-waves is not
sufficiently small and may be necessary to consider second order terms in the expansion of $\coth\left(\frac{\beta\omega_{\bf k}}{2}\right)$.
We will use the classical approximation only for a qualitative analysis at low temperatures while the correct results will be directly
determined by Eq. (\ref{eq:rho_scha}).
\begin{figure}[h]
\centering
\epsfig{file=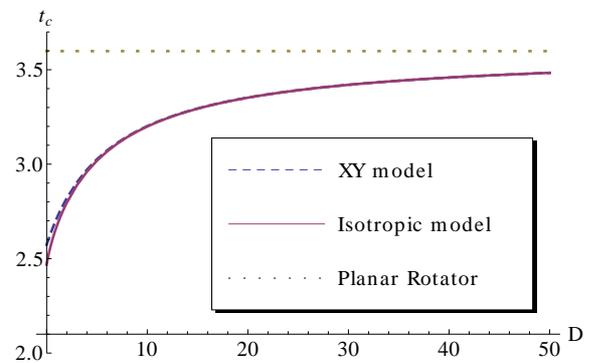,width=\linewidth}
\caption{The reduced critical temperature $t_c=\frac{T_c}{J\tilde{S}^2}$ as function of $D$ (in units of $J$) obtained from classical
approximation given by Eq. (\ref{eq:tc}) for $p=3$.}
\label{fig:tc_classical}
\end{figure}

The critical temperature is obtained at the point where Eq. (\ref{eq:rho_cl}) admits as solution the trivial one. Considering only
first order terms, $T_c$ is determined by
\begin{eqnarray}
\label{eq:tc}
\frac{T_c}{J\tilde{S}^2}=\frac{1}{e R+I}
\end{eqnarray}
with $e$ being the base of the natural logarithm. The classical result for $T_c$ shows an unexpected increasing behavior with
respect to $D$. In the large-$D$ phase, there is no long-range order, even at zero temperature; however the classical model indicates
a finite transition temperature. The problem is to consider low-energy spin-waves when quantum fluctuations are larger enough to
disorder the system. Classically, the limit $D\to\infty$ provides the transition temperature for the planar rotator once there is no
$S^z$ component. In Fig. (\ref{fig:tc_classical}) we plot $T_c$ as a function of $D$ using the classical approximation for $p=3$. The
curve for the $XY$ model ($\lambda=0$), the isotropic model ($\lambda=1$) as well as the asymptotic limit of the planar rotator are
plotted. The $\lambda$ influence is small, mainly for $D\gg 1$, and both curves tend to the planar rotator limit $T_c=3.60 J\tilde{S}^2$
when $D$ increases. A similar behavior is observed for all values of the power-law exponent $p$.

\begin{figure}[h]
\centering
\epsfig{file=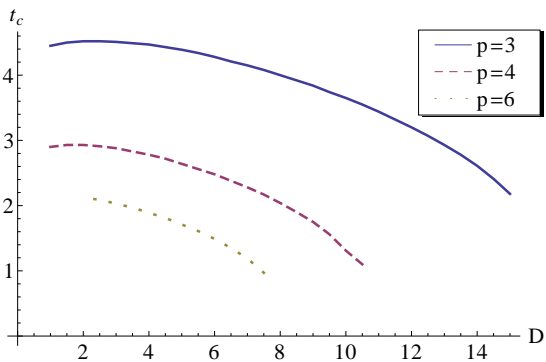,width=\linewidth}
\caption{The reduced critical temperature $t_c$ as function of $D$ using the Eq. (\ref{eq:rho_scha}). At $D=D_c$ a QFT occurs and for
$D>D_c$ there is no finite transition temperature. $D$ is given in units of $J$.}
\label{fig:Tc_D}
\end{figure}

To include quantum fluctuations in the critical temperature, we have to evaluate the phase transition using the Eq. (\ref{eq:rho_scha}),
disregarding any classical approximation. The results of $T_c$ as a function of the single-ion anisotropy $D$ for $p=$3, $4$ and $6$ are shown
in Fig. (\ref{fig:Tc_D}). As expected, $T_c$ decreases as $D$ increases and close to the critical point $D_c$, there is a discontinuity
associated with the $QFT$. Above the critical anisotropy there is no transition at finite temperature and the system is quantum disordered.
The according results between classical and quantum calculations are recovered in the limit of large spin in which fluctuations are negligible.

\begin{figure}[h]
\centering
\epsfig{file=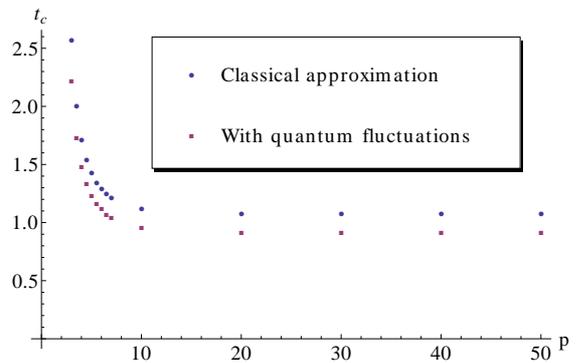,width=\linewidth}
\caption{The reduced critical temperature $t_c$ as function of the power-law exponent $p$ for $D=0$.}
\label{fig:tc_p}
\end{figure}

The long-range interactions effects are shown in Fig. (\ref{fig:tc_p}). We plot the results for the $XY$ model (with $D=0$) taking into
account the quantum fluctuations, through Eq. (\ref{eq:rho_scha}), as well as using the classical approximation from Eq. (\ref{eq:rho_cl}).
For $p=3$ we have found $T_c=2.57 J\tilde{S}^2$ by using the classical approximation and $T_c=2.22 J\tilde{S}^2$ considering the
spin-waves contributions. The large values for $T_c$, compared with the results for $p\gg 1$, reflect the large exchange spin
energy and so the high thermal energy required for restoring the $O(3)$ symmetry. On the other hand, in the limit $p\gg 1$, the critical temperature
tends to $T_c=1.08J\tilde{S}^2$ (classical) and $T_c=0.91J\tilde{S}^2$ (quantum). The result is consistent with the transition
temperature $T_c=1.076 J\tilde{S}^2$ obtained for the classical $XY$ model with nearest-neighbor interactions \cite{PRB48-b}.
Furthermore, the classical approximation overestimates the critical temperature which justifies the smaller temperature determined
from Eq. (\ref{eq:rho_scha}).

\begin{figure}[h]
\centering
\epsfig{file=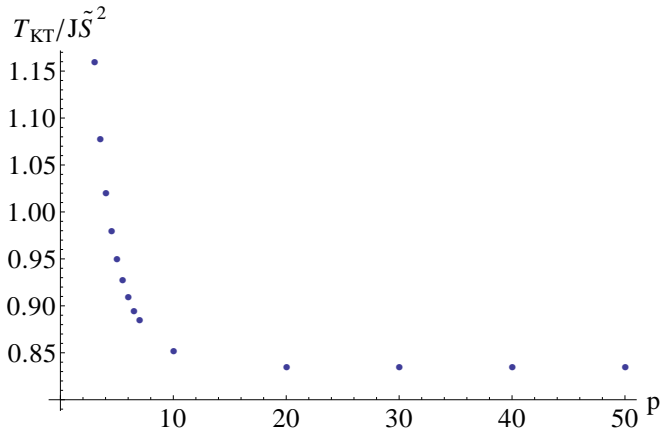,width=\linewidth}
\caption{The temperature $T_{KT}$ as function of the power-law exponent $p$ for $\lambda=0$.}
\label{fig:tkt}
\end{figure}

It is well-known that the stiffness for the $XY$ model with short-range interactions should exhibit a universal jump at $T_{KT}$
associated with the Berezinskii-Kosterlitz-Thouless ($BKT$) transition. The $BKT$ transition is not associated with a spontaneous symmetry-breaking
as occurs in the transitions described by the Mermin-Wagner theorem. Instead, it involves the emergence of topological configurations
at finite temperature, the magnetic vortices. The angle field $\phi({\bf r})$ should be split in two parts
$\phi({\bf r})=\phi_0({\bf r})+\psi({\bf r})$ where $\phi_0({\bf r})$  describes the small phase fluctuations of the order parameter
while $\psi({\bf r})$ is the vortex field. The stiffness $\rho_s$ from Eq. (\ref{eq:rho}) takes into account only the spin-wave contribution,
neglecting any vortex effect. The vortex field can be implemented by the replacement $\rho_s\to\tilde{\rho}_s=\rho_s\eta$, in which
$\eta$ is a renormalization factor. The $\eta$ term presents a discontinuity at $T_{KT}$ temperature given by
$\lim_{T\to T_{KT}}J\tilde{S}^2\eta/T_{KT}=2/\pi$. Below $T_{KT}$, an ordered state occurs due the bounding of the vortex-antivortex
pairs whilst above $T_{KT}$, the vortices are free guiding the system to a disordered state. The critical temperature $T_c$
obtained from Eq. (\ref{eq:tc}) is only a first approximation to $T_{KT}$ while a better result is determined by the crossing between
the curve $\rho_s$, from Eq. (\ref{eq:rho_cl}), and the line $\eta=2T/\pi J\tilde{S}^2$ \cite{EPJB44, PRB48-b}. In general, the
$T_{KT}$ temperature is smaller than $T_c$ and for $T<T_c$, the $SCHA$ provides good results at finite temperatures. For the
two-dimensional $XY$ model, for example, $T_{KT}=0.83 J\tilde{S}^2$ \cite{PRB48-b}. Applying the same procedure to the $XY$
ferromagnet model with long-range interactions, we have found $T_{KT}=1.16 J\tilde{S}^2$ with $p=3$ and $T_{KT}=0.83 J\tilde{S}^2$
in the limit $p\gg 1$. Figure (\ref{fig:tkt}) shows $T_{KT}$ as a function of $p$ for the $XY$ model.

\section{Conclusion}
\label{conclusion}
In this work we have investigated the two-dimensional single-ion anisotropic ferromagnetic with long-range interactions using a
Schwinger bosonic formalism and the Self-consistent Harmonic Approximation. As in the short-range interaction case,
the single-ion anisotropy $D$ is responsible for a quantum phase transition. The Hamiltonian is composed by two conflicting terms:
the exchange and the single-ion anisotropic ones. Below a critical value $D_c$ the system has long-range order once an ordered state
is energetic favorable. On the other hand, for the large-$D$ phase ($D>D_c$), there is a high cost to keep out-of-plane spin components and,
therefore, the system prefers to keep a vanishing spontaneous magnetization state (even at zero temperature). The results obtained
from both methods are similar and for large power-law exponent $p$ there is an excellent according with short-range interaction
models. It is expected since for $p\gg 1$, the long-range interactions are very weak at long distances. Actually, in the approximation
for long wavelength ($k\ll 2\pi/a$), the long-rage interaction model with $p>4$ has a relativistic energy spectrum, as occurs for a system
with only short-range interactions. For small values of $p$, the critical anisotropy $D_c$ is larger than that of the short-range version, which
reflects the high energetic cost to flip spins.

At finite low temperatures, there is a thermal phase transition for $D<D_c$. Using the $SCHA$, we have determined the critical temperature
$T_c$ which separates a state with quasi long-range order from another one with vanishing short spin-spin correlation. In analogy with
nearest-neighbor interaction models, there is a temperature $T_{KT}$ associated with the $BKT$ transition. The $SCHA$ does not consider vortex
effects but they can be introduced by a renormalization factor in $\rho_s$. The critical $T_c$ gives only a first approximation to $T_{KT}$ and
more precise results are obtained by the intersection of the stiffness $\rho_s$ curve with  the line $\eta=2T/\pi J\tilde{S}^2$. Considering
$D\approx 0$ and $p\gg 1$ we can use a classical approximation at sufficient low finite temperatures. However, for $D\approx D_c$ the quantum
fluctuations are large and the approximation fails. While the classical result for $D\to\infty$ predicts a finite temperature $T_c$ (below
which there is a ordered state), the quantum model indicates a disordered state even at zero temperature for any $D>D_c$. For large spins,
the spin-wave fluctuations are negligible and we have according results for both limits, quantum and classical. As expected, the limit $p\gg 1$
always provides comparable results with the well-known short-range interaction models.

\bibliography{manuscript.bib}
\end{document}